# SafeLight: Enhancing Security in Optical Convolutional Neural Network Accelerators

Salma Afifi
Department of Electrical and Computer Engineering
Colorado State University
Fort Collins, USA
salma.afifi@colostate.edu

Ishan Thakkar
Department of Electrical and Computer Engineering
University of Kentucky
Lexington, USA
igthakkar@uky.edu

Sudeep Pasricha
Department of Electrical and Computer Engineering
Colorado State University
Fort Collins, USA
sudeep@colostate.edu

*Abstract*— The rapid proliferation of deep learning has revolutionized computing hardware, driving innovations to improve computationally expensive multiply-and-accumulate operations in deep neural networks. Among these innovations are integrated silicon-photonic systems that have emerged as energy-efficient platforms capable of achieving light speed computation and communication, positioning optical neural network (ONN) platforms as a transformative technology for accelerating deep learning models such as convolutional neural networks (CNNs). However, the increasing complexity of optical hardware introduces new vulnerabilities, notably the risk of hardware trojan (HT) attacks. Despite the growing interest in ONN platforms, little attention has been given to how HT-induced threats can compromise performance and security. This paper presents an in-depth analysis of the impact of such attacks on the performance of CNN models accelerated by ONN accelerators. Specifically, we show how HTs can compromise microring resonators (MRs) in a state-of-the-art non-coherent ONN accelerator and reduce classification accuracy across CNN models by up to 7.49% to 80.46% by just targeting 10% of MRs. We then propose techniques to enhance ONN accelerator robustness against these attacks and show how the best techniques can effectively recover the accuracy drops.

## I. INTRODUCTION

In recent years, a wave of advanced applications, such as autonomous vehicles, intelligent robotics, fake news detection, and pandemic trend forecasting, has been powered by sophisticated machine learning models. As researchers continue to develop increasingly large and deep neural network (DNN) models like multi-layer perceptrons (MLPs) and convolutional neural networks (CNNs), the demand for hardware platforms that can deliver superior performance while meeting strict power efficiency requirements has grown. To address this need, specialized hardware accelerators such as Google's TPU and Amazon's Inferentia have been designed, offering significantly improved performance-per-watt for DNN execution compared to CPUs and GPUs [1].

However, electronic accelerator architectures are now facing inherent limitations in the post-Moore era due to high fabrication costs and diminishing performance improvements with technology scaling [2]-[4]. A significant challenge lies in the electronic transmission of data over metallic wires, which creates bottlenecks in both bandwidth and energy consumption. As a solution, silicon photonics has emerged as a groundbreaking technology, offering ultra-high bandwidth, low latency, and energy-efficient data communication. Silicon photonics has already replaced traditional metallic interconnects for high-speed data transfers and is now being actively integrated at the chip level [5], [6]. Furthermore, silicon photonics has the potential to accelerate computations in DNNs. By using optical components for operations such as matrix-vector multiplication, it has become possible to develop a new generation of hardware accelerators for deep learning [7]. These accelerators, built with photonic integrated circuits (PICs) featuring on-chip waveguides and CMOS-compatible electro-optic devices facilitate low-latency and energy-efficient data processing in the optical domain.

Despite the significant advantages and increasing adoption of silicon photonics in industry and research, the rising complexity of silicon-photonic hardware introduces new vulnerabilities, notably the risk of hardware trojan (HT) induced attacks. The malicious alterations with HTs can significantly compromise the functionality, performance, and security of optical neural networks (ONNs). This paper investigates these largely unexamined security threats in photonic accelerators, providing a thorough analysis of their impact on CNN inference performance. For the first time, we present an in-depth susceptibility analysis of various HT attacks targeting non-coherent optical CNN hardware accelerators, specifically focusing on the operation of microring resonator (MR) devices. Additionally, we propose and evaluate software-based attack mitigation techniques, demonstrating their effectiveness in significantly enhancing the robustness of CNN models against these emerging threats in ONNs. The key novel contributions of this paper are:

- A comprehensive exploration of HT attacks on non-coherent ONN accelerators for CNNs, with a focus on attacks that compromise the operation of MRs, which are key electro-optic devices required for high-fidelity computations in non-coherent ONN accelerators;
- A susceptibility analysis examining the effects of HT-induced actuation and thermal hotspot attacks in ONN hardware on the performance of various CNN models;
- An extensive evaluation of software-based attack mitigation strategies to identify the most effective ones that can help overcome the impact of HT attacks in ONN accelerators during CNN model execution.

## II. BACKGROUND

### A. Deep Neural Networks and Security Threats

As machine learning becomes deeply integrated into various aspects of daily life and the size of DNNs continues to grow, security concerns around these models are escalating. A notable software-based threat is adversarial attacks, where maliciously crafted inputs (e.g., images), though perceptually similar to legitimate ones are designed to deceive learning models [8]. Another class of security threats are trojan attacks which fall into two categories: Model Trojans (MT) and Hardware Trojans (HT). MTs exploit inherent vulnerabilities of the neural network model, where an attacker manipulates certain weights to cause the network to malfunction when triggered. HTs involve malicious circuits with a trigger and a payload; the payload activates when the trigger condition is met [9]. The authors in [10] demonstrate DNN misclassification attacks by using HTs for fault injection in

SRAM and DRAM to alter bit values, assuming detailed knowledge of the model and hardware. Li et al. [11] embed HT circuits to create malicious DNN models, necessitating retraining to preserve original accuracy. In [12], malicious behavior is injected during DNN execution by modifying multiplexer logic and internal operations. Although various mitigation techniques, such as adversarial training [8], have been explored to address these security threats, no attention has been given to mitigating such threats in ONN platforms.

*B. Silicon Photonics for DNN Hardware Acceleration*

Optical accelerators for DNNs have gained considerable attention due to their promising performance and energy efficiency [7], [13]-[17]. These accelerators target accelerating multiply-and-accumulate (MAC) operations and can be categorized into coherent and non-coherent architectures. Coherent architectures encode DNN parameters in the phase of the optical signal. Non-coherent architectures employ multiple wavelengths and DNN parameters are encoded in the amplitude of the optical signal, allowing for parallel processing of operations across different wavelengths [5].

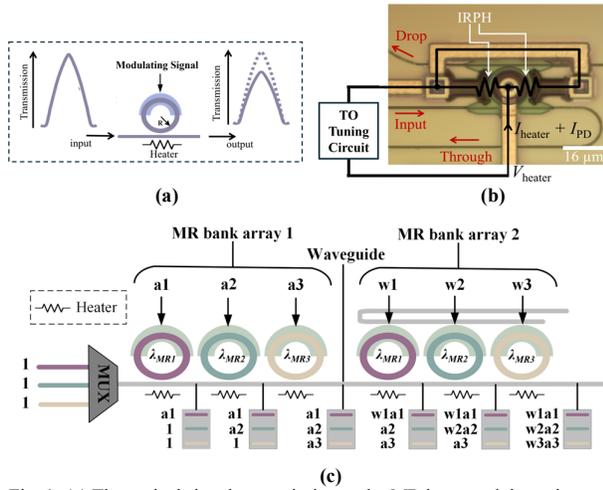

Fig. 1: (a) The optical signal transmission at the MR input and through ports' before and after imprinting a DNN parameter (e.g., weight); (b) Microscope image of an MR overlaid with an in-resonator photoconductive heater (IRPH) used in TO tuning [21]; (c) MR bank arrays used to perform multiplication by imprinting input vector (a1-a3), followed by weight vector (w1-w3).

In non-coherent ONN accelerators, MRs are fundamental optoelectronic devices used for executing critical operations (see Fig. 1(a)). Each MR is precisely tuned to operate at a specific resonant wavelength ($\lambda_{MR}$), which is defined by:

$$\lambda_{MR} = \frac{2\pi R}{m} n_{eff}, \qquad (1)$$

where $R$ is the MR radius, $m$ denotes the resonance order, and $n_{eff}$ represents the effective index of the device. Adjusting $n_{eff}$ to affect MR resonance is typically enabled through two types of peripheral circuits: *(i)* a signal modulation circuit that enables modulation of electronic data onto an optical signal, *(ii)* a tuning circuit that biases MR resonance to counter resonance shifts induced due to fabrication and thermal variations. These peripheral circuits can be thermo-optic (TO) [18] or carrier injection electro-optic (EO) [19] or combined EO-TO [24]. The EO method is quicker (≈ns range) and consumes less power (≈4 µW/nm); however, it cannot be used for large tuning ranges [18]. Conversely, the TO method offers a larger tunability range, but has higher latency (≈µs range) and power consumption (≈27 mW/FSR) [19].

To enhance throughput and replicate neural functions in DNNs, non-coherent architectures employ wavelength-division multiplexing (WDM). This technique involves combining multiple optical signals with distinct wavelengths into a single waveguide via an optical multiplexer [20]. The waveguide then traverses a bank of MRs, each set to a specific wavelength, allowing simultaneous execution of multiple multiplications. For example, as illustrated in Fig. 1(c), the process of multiplying an input vector $[a_1, a_2, a_3]$ by a weight vector $[w_1, w_2, w_3]$ is achieved using two MR bank arrays: the first array encodes input activations, while the second performs the multiplications. The resulting dot product is computed by summing the three optical signals (summing not shown in the figure).

*C. Attacks in Optical Hardware Accelerators*

Analyzing the inference process in ONN accelerators reveals a broad array of potential attack surfaces within the hardware. Fig. 2 provides an overview of the fundamental devices and circuits utilized in non-coherent optical computing with those particularly more vulnerable to HTs highlighted via red boxes. Through various attack vectors, if an attacker gains access to any of these devices, it can severely compromise the security, functionality, and performance of the accelerated model. While previous research has not explored the impact of such attacks on the overall operation of ONN accelerators, several studies have investigated threats to specific components used in non-coherent silicon photonic accelerators. For instance, [9] explored memory attacks in electronic DNN accelerators. They proposed an image-triggering method to activate HT with specific input images. The HT, once triggered, degrades the model's accuracy.

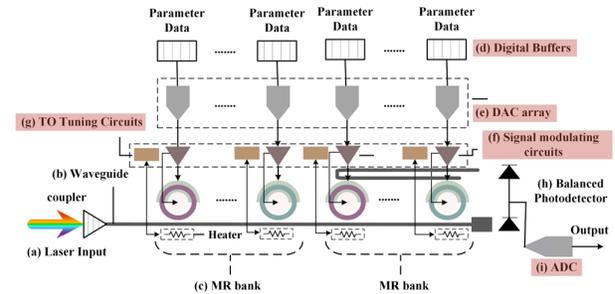

Fig. 2: Overview of photonic circuit used to implement operations for DNN acceleration, composed of (a) a laser source; (b) a waveguide; (c) MR banks to perform MAC operations; (d) the banks are tuned as per data from the electronic buffers; (e) using DACs; (f) to tune the MR devices; (g) the result from the MAC operation is detected and accumulated using a photodetector (PD); (f) for converting the data to digital domain, an ADC is used.

HTs in analog circuits are usually uncommon due to the high sensitivity of analog paths which usually results in easy detection of such malicious circuits. Nevertheless, [22] details how analog circuits, such as digital-to-analog converters (DACs) can still be attacked by utilizing the chip's test bus. Analog-to-digital (ADCs) converters required for tuning the MR devices with the correct DNN model parameters can also be attacked using methods like [23]. As shown in Fig. 2, the ADC converts the final partial sum of the dot product computed in a row of MR banks. Accordingly, attacking the ADCs in an ONN accelerator would impact and change several outputs during DNN execution and can result in significant accuracy losses at inference time. The tuning circuits also pose a significant threat as an attack vector due to their direct influence on MR devices. HTs can be embedded within these circuits to disrupt the operation of MRs [24], e.g., an HT can force an MR into "off-resonance" state, preventing it from tuning to the desired wavelength.

Previous work has also explored defense mechanisms against snooping attacks in silicon-photonic-based systems. SOTERIA [25] presents a framework that utilizes process variation-based authentication signatures combined with architecture-level enhancements to protect data in photonic network-on-chip (PNoC) architectures from snooping attacks. Another framework was proposed in [26] to improve hardware security in optoelectronic systems by using cryptographic keys derived from optical lithography imperfections and an online detection mechanism. *To the best of our knowledge, no prior work has explored HT attacks or defense mechanisms for non-coherent ONN accelerators.*

### III. ATTACKS ON OPTICAL CNN ACCELERATORS

#### A. Optical Hardware Accelerator Architecture

Fig. 3 presents an overview of a state-of-the-art optical CNN accelerator leveraged in our analysis and based on the Crosslight architecture from [7]. The photonic substrate utilizes MR devices to perform vector dot product (VDP) operations and employs optoelectronic PDs for summing results across multiple wavelengths. An electronic control unit manages the photonic devices, facilitates communication with global memory to retrieve parameter values, maps vectors, and handles partial sum buffering. DAC arrays are used to transform buffered signals into analog tuning signals for the MRs. Meanwhile, ADC arrays convert the analog signals produced by the PDs into digital values, which are then sent back for post-processing and buffering. For optimal performance, the photonic substrate responsible for performing the optical VDP operations is divided into two main sections: a CONV block responsible for accelerating the convolution layers and an FC block for accelerating the fully connected layers. For each block, the partial sums computed by each VDP unit at any time are accumulated and summed using an optical summation block.

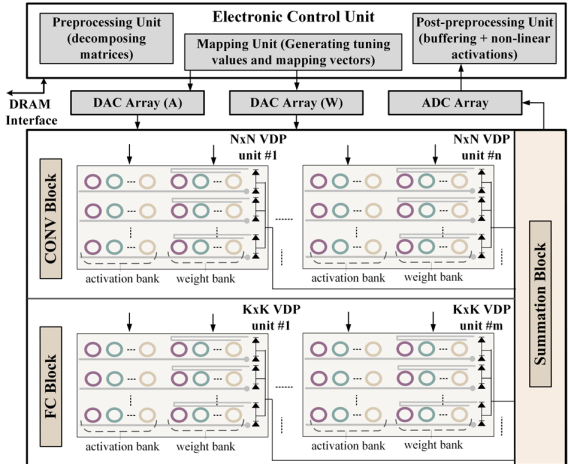

Fig. 3: Optical CNN accelerator architecture

#### B. Optical Accelerator Attack Analysis

As discussed in section II.C, ONN accelerators are vulnerable to HT attacks that target optical components. Based on the CNN accelerator architecture shown in Fig. 3, the attack surfaces in our optical hardware accelerator include the convolution (CONV) and the fully connected (FC) blocks, where any of the MR banks may be compromised. This paper focuses on the operation of MR devices, which are critical for imprinting model parameters (weights) and inputs (activations) onto optical waveguides for computation. The attack vectors in our framework target the MR peripheral circuits, where HTs in the signal modulation (actuation) and tuning control units (see Fig. 1(b) and Fig. 2) can disrupt the MR operation. In the following subsection, we present the two attack vectors considered in our analysis.

*1) MR Actuation Attacks*

Given the large number of MR devices present in an ONN accelerator such as [7], we assume an attacker can introduce HTs randomly within the accelerator substrate to manipulate a small portion of these MRs. Such HTs could specifically target the EO circuits used in the signals' actuation. In this scenario, each HT circuit would interfere with a single MR, causing it to enter an "off-resonance" state, where it is no longer tuned to function at the intended wavelength. Fig. 4 illustrates the operation of an MR bank array in the event of a successful attack on the second MR device in the first MR bank array, originally tuned to operate on $\lambda_{MR2}$. Using the same example in Fig. 1(c), the MR bank arrays were initially expected to compute the dot product of the input vector $[a_1, a_2, a_3]$ and weight vector $[w_1, w_2, w_3]$. Following the attack on the MR highlighted in red in Fig. 4, the second term in the dot product becomes corrupted (shown in red).

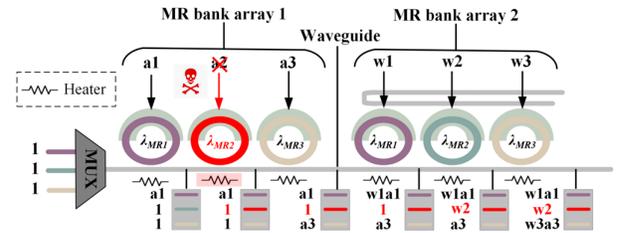

Fig. 4: Actuation attack on an MR bank used to perform multiplications.

*2) Thermal Hotspot Attacks*

The second attack vector involves thermal-based hotspot attacks. Typically, the MR banks in a silicon-photonic chip are designed to resonate with and operate upon their assigned carrier wavelengths at a specific temperature [24]. The tuning circuit is usually designed to manage minor temperature fluctuations caused by workload variations, ensuring the MR continues to function correctly under such conditions. However, if the temperature exceeds a certain threshold, it can cause a shift in the MR's resonance wavelength. Typically, an optical waveguide is configured to support a specific number of evenly spaced wavelengths, corresponding to the number of columns in each MR bank [20]. Considering the example presented in Fig. 1(c), the accommodated wavelengths are $\lambda_{MR1}, \lambda_{MR2}$, and $\lambda_{MR3}$. In the event of a large temperature change, the resonance wavelength of each MR shifts away from its assigned carrier wavelength. Considering a thermal attack focused on the three MRs in a bank as shown in Fig. 5, the first MR will operate on an unsupported wavelength, and the second and third MRs will now operate on wavelengths $\lambda_{MR1}$ and $\lambda_{MR2}$ respectively. This will accordingly result in corrupted output values for all of the multiplications as shown (in red).

The resonance shift $\Delta\lambda_{MR}$, due to a temperature change can be modeled using the following equation from [20]:

$$\Delta\lambda_{MR} = \Gamma_{Si} \cdot \frac{\delta n_{Si}}{\delta T} \cdot \frac{\lambda_{MR}}{n_g} \cdot \Delta T, \qquad (2)$$

where $n_g$ is the group refractive index (ratio of speed of light to group velocity of all wavelengths in the waveguide) of the MR waveguide, $\Gamma_{Si}$ is the modal confinement factors of the MR's core (Si), and $\frac{\delta n_{Si}}{\delta T}$ is the thermo-optic coefficient of Si.

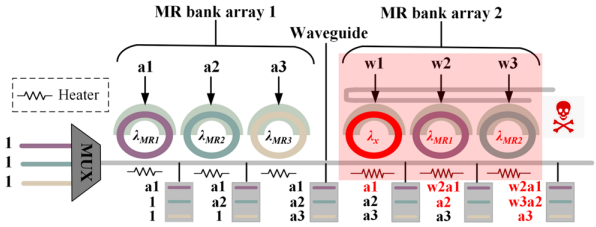

Fig. 5: Thermal hotspot attack on an MR bank used to perform multiplications.

Fig. 6 shows two thermal hotspot attacks on the CONV block of the ONN hardware accelerator illustrated in Fig. 3 where heaters across groups of MRs in banks are manipulated using HTs. This can be executed using a HT that specifically targets the TO tuning circuit (see Fig. 1(b)). By attacking a single heater, a hotspot attack can be triggered, leading to a localized temperature increase in one MR, which subsequently impacts the surrounding chip area and neighboring MRs. The heatmap, generated using the HotSpot tool [27], simulates an attack scenario where two MR banks have multiple compromised heaters that cause higher heat dissipation. The temperature rise from this heat, influenced by the MR bank's location and the severity of the thermal hotspot attack, affects not only the targeted CONV MR bank arrays but also neighboring banks.

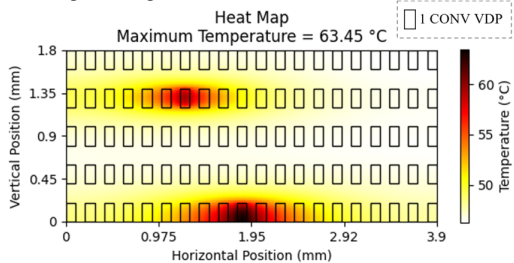

Fig. 6: Heatmap for the CONV MR bank arrays with hotspot attacks.

## IV. ONN ACCELERATOR ATTACK SUSCEPTIBILITY ANALYSIS

This section presents susceptibility analysis conducted to quantify the impact of the attack vectors described in Section III on the performance of CNN models executing on the compromised ONN accelerator. We evaluate three CNN models in our analysis: a simple MNIST-classifier model, a ResNet18 model, and a large VGG16 variant with six CONV layers. The parameters for each model are shown in Table I.

TABLE I: CNN MODELS PARAMETERS

|  | CNN_1 | ResNet18 | VGG16_v |
|---|---|---|---|
| Dataset | MNIST | CIFAR10 | Imagenette |
| CONV Layers | 2 | 17 | 6 |
| CONV Parameters | 2.6K | 4.7M | 3.9M |
| FC Layers | 3 | 1 | 3 |
| FC Parameters | 41.6K | 5.1K | 119.6M |
| Total Parameters | 44.2K | 4.7M | 123.5M |

To assess the impact of the HT attacks on the ONN accelerator from [7], we developed a comprehensive Python simulator that modeled the accelerator based on device, circuit, and component level characteristics defined in [7], with the ability to introduce HT attacks in the optical accelerator during model inference. This was achieved by modifying the models' parameters based on their mapping to the ONN accelerator. All layers of the models (as specified in Table 1) were mapped using a weight-stationary approach. The CONV block in the ONN accelerator is composed of $m = 100$ VDP units where each unit is comprised of $20 \times 20$ MRs. The FC block is composed of $n = 60$ VDP units where each unit contains $150 \times 150$ MRs.

For the two attack vectors examined in this paper—actuation attacks and thermal hotspot attacks on MRs—we analyze a total of nine scenarios. In the first set of three cases, 1%, 5%, and 10% of the MRs in the CONV block are targeted with HT attacks (actuation and hotspot). Next, we consider attacks on 1%, 5%, and 10% of the MR banks in the FC block. Lastly, the same analysis is performed across the entire ONN accelerator architecture (CONV + FC), with 1%, 5%, and 10% of the total MRs under attack.

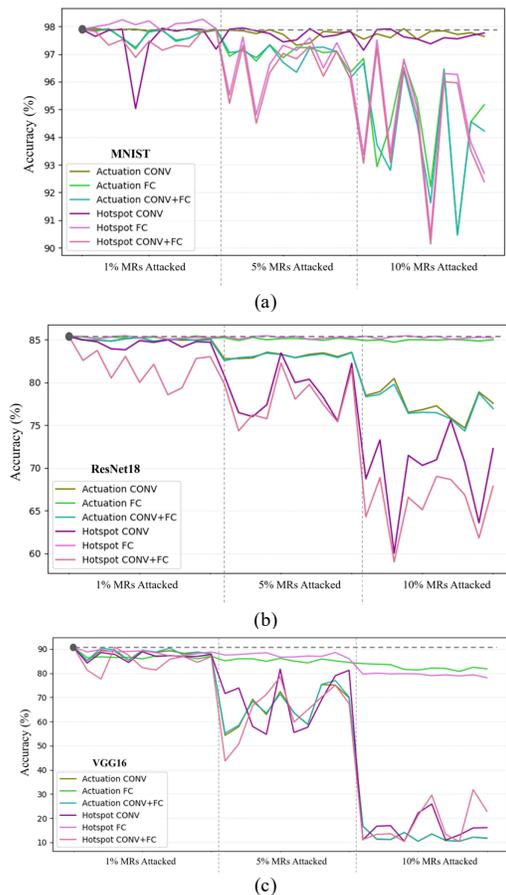

Fig. 7: Impact of actuation and hotspot attacks on (a) MNIST-based CNN model; (b) ResNet18 CNN model; (c) a VGG16 variant CNN model.

Fig. 7(a)-(c) illustrate the accuracies of the three CNN models in the presence of the actuation and hotspot attacks, with different attack intensities. For both the actuation and hotspot attacks, we simulate 10 different uniformly distributed combinations of individual random MRs (for actuation attacks) and banks of MRs (for hotspot attacks) being targeted by HTs for the 1%, 5%, and 10% cases (the x-axes of the figures are divided into 3 groups of 10 results).

As illustrated in Fig. 7, the susceptibility of each model to MR attacks varies based on its architecture. For instance, in the MNIST model, attacking the FC block leads to more significant accuracy drops. In contrast, for ResNet18, where the CONV layers are more prominent than the FC layer, accuracy declines are more pronounced when the CONV block is targeted. Similarly, in the VGG16 variant model, despite having a greater total number of parameters in the FC layers, the higher importance of the CONV layers results in more pronounced accuracy declines when the CONV block is attacked. In fact, in the cases of attacking 10% of the MRs in the CONV and FC blocks with the VGG16 model, the model becomes almost completely corrupted as shown in Fig. 7(c). This is due to the massive size of VGG16 model which requires mutliple mappings for each layer onto the ONN accelerator. Across all models, thermal hotspot attacks consistently result in greater accuracy losses compared to actuation attacks. Although both attack types affect the same number of MRs, thermal hotspots corrupt clusters of model parameters, which our analysis shows can have a more substantial impact. Specifically, with the worst attack case of hotspot attacks in both CONV and FC blocks affecting 10% of the total MRs available, MNIST, ResNet18, and VGG16-variant models experienced accuracy drops of 7.49%, 26.4%, and 80.46%.

*In summary*, the key insights from our susceptibility analysis are: *1)* HTs, whether randomly distributed to create actuation attacks across individual MRs, or orchestrating thermal attacks in groups of MRs, cause significant accuracy degradation in CNNs executing on ONN accelerators, *2)* larger CNN models, which map parameters from multiple layers to shared VDP blocks, experience a steeper accuracy decline from HT attacks, *3)* a model's susceptibility is influenced by the composition of its FC and CONV layers, and *4)* hotspot attacks are more potent than actuation attacks.

## V. HT Attack Mitigation Techniques

As discussed in the previous section, the impact of attacks from distributed HTs in ONN accelerators can be very potent. Mitigating all potential HT circuits at the hardware level is a cumbersome challenge. Hardware-based solutions can significantly increase design complexity, cost, and power overheads, and they often lack flexibility to adapt to new threats, variations in fabrication processes, or architectural changes. In contrast, software-based mitigation techniques offer a more scalable, adaptable, and cost-effective approach to enhancing ONN accelerator resilience to attacks. Accordingly, we focus on software-based mitigation techniques. This is the first study to explore software-based solutions with the goal of making ONN accelerators more robust against HT attacks, as discussed next.

### A. Regularization techniques

Regularization techniques in CNNs have been widely explored and proposed as solutions to overfitting [28]. In particular, the L2 regularization technique has been widely adopted due to its effectiveness in increasing the model's robustness and resulting in better predictable accuracy [29]. In the L2 regularization technique, the sum of the squared parameters of a given model is multiplied by a hyperparameter and added to the loss function as the penalty term to be minimized. This regularization strategy shifts the weights closer to the origin by adding the regularization term $R(w) = \frac{\lambda}{2m} * \Sigma \|w\|^2$ to the cost function, where $\lambda$ is the regularization parameter, and $m$ is the number of samples [30]. In the context of HT attacks on ONN accelerators, the lower accuracies after an attack can be attributed to the change in the relative strengths of the output neurons in the presence of noise from the attacks [31]. By reducing the variance of these neuron values during training using L2 regularization, for the first time, we explore how to preserve the relative strength of neurons and thereby reduce the degradation of accuracy even in the presence of multiple HTs in the ONN MR banks.

### B. Noise-aware training

Noise-aware training has demonstrated its effectiveness in producing models that are able to generalize better and reduce susceptibility to device noise, making it particularly valuable for emerging accelerator technologies like PCM-based AI accelerators with noise-prone components [32], [33]. For the first time, we explore how to utilize this strategy for ONN accelerators, to reduce the degradation of accuracy in the presence of multiple HTs in the ONN MR banks. The intuition is that attacks that corrupt model parameters can be considered a form of "noise"; however, unlike other noise sources, the unpredictable nature of such attacks makes identifying the exact noise pattern challenging. In our approach, we demonstrate that introducing random Gaussian noise into model layers during training enhances the robustness of ONN accelerators to various HT attacks.

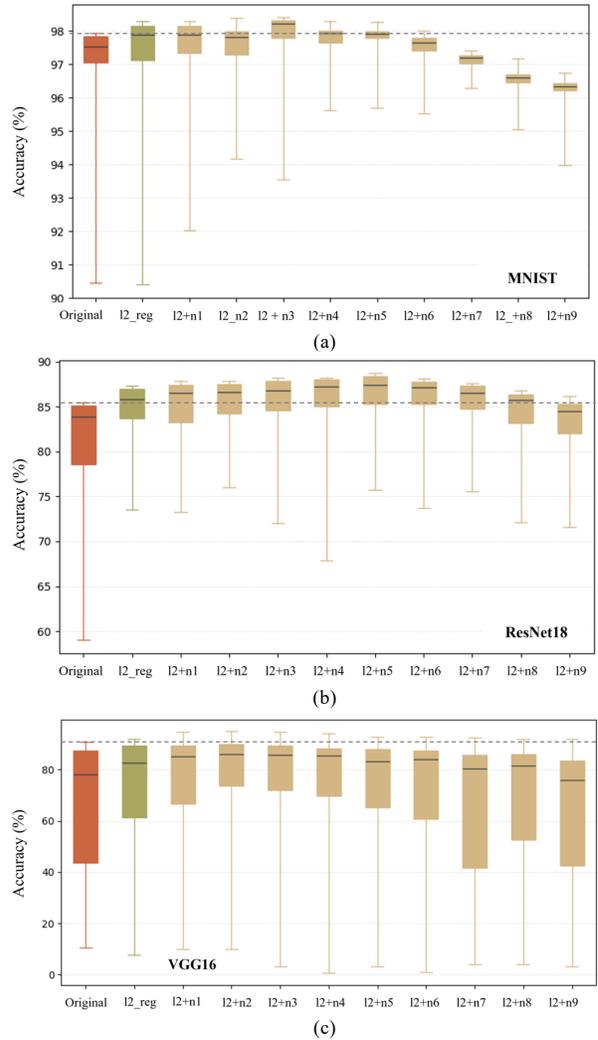

Fig. 8: Accuracies obtained with different model variants for (a) MNIST-based model, (b) ResNet18 model, and (c) VGG16 variant model

## VI. ONN Accelerator Attack Mitigation Analysis

To investigate the impact of the two software-based HT attack mitigation approaches, we utilized our simulator with

HTs introduced into the ONN accelerator, for the case of inference execution of the three CNN models described in Section IV. The models were initially trained in PyTorch, generating nine variants through noise-aware training, with Gaussian noise standard deviations ranging from 0.1 to 0.9. Additionally, the models were trained both with and without L2 regularization. Variants that incorporated both L2 regularization and noise-aware training were also considered.

We first conducted an in-depth analysis to evaluate the effectiveness of each model variant under different HT attack scenarios. The focus was on assessing the performance of noise-resilient models trained with Gaussian noise at different standard deviations, as well as models trained with L2 regularization alone and in combination with noise-aware training. Fig. 8 shows a box and whiskers plot that summarizes (for each model variant on the x-axis) the accuracy results across all attack scenarios (1%, 5%, and 10% actuation attacks and thermal hotspot attacks) targeting the MRs in the CONV and FC blocks of the ONN accelerator. The CNN model variants compared along the x-axis include the original model without any enhancements (*Original*), the model with only L2 regularization (*L2_reg*), and models with both L2 regularization and noise-aware training, using Gaussian noise with standard deviations between 0.1 and 0.9 (*l2+n1 – l2+n9*).

The baseline accuracy for each model is shown by the horizontal lines in the figures. As observed, the robust models not only recovered a significant portion of the accuracy degradations due to the various attacks, but they also surpassed the baseline accuracy values due to their improved generalization. The results indicate that combining L2 regularization with Gaussian noise-aware training consistently enhances robustness, as seen in the higher accuracy values and smaller accuracy variations compared to the original models. However, different models achieve optimal robustness with different noise standard deviations. Specifically, the most robust configurations identified were l2+n3 for the MNIST-based model, l2+n5 for ResNet18, and l2+n2 for the VGG16 variant.

Fig. 9 illustrates a detailed performance comparison between the most robust model variants across the three CNN models (from the analysis in Fig. 8) and the original models under actuation and thermal hotspot attacks affecting 1%, 5%, and 10% of the total MRs in the ONN accelerator. The baseline accuracy for each original model is shown by the horizontal dashed lines in the figures. The robust models consistently demonstrate superior accuracy compared to the original models, with significantly higher resilience to the corruption introduced by HT attacks. In all scenarios, as the number of attacked MRs increases, the robust models exhibit a much smaller accuracy drop compared to the original models. The only exception is in the case of attacking 10% of the MRs with the VGG16 model where the model becomes significantly corrupted with accuracy degradations reaching 80.46%, where significant recovery is difficult.

In the most severe case of attacks affecting 10% of the MRs (with hotspot attacks), the MNIST, ResNet18, and VGG16 original models experience an accuracy degradation of up to 7.49%, 26.4%, and 80.46%. For MNIST, ResNet18, and VGG16 models, the proposed robust models are able to recover up to 5.4%, 21.2%, and 30.7% of the accuracy drops. However, a scenario with a 10% attack is less likely than ones targeting 1% to 5% of the MRs, which involve fewer HTs that would be harder to detect. For the most severe case of attacks affecting 5% of the MRs (with hotspot attacks), the MNIST, ResNet18 and VGG16 models experience accuracy drops up to 3.39%, 11.1%, and 47.2%. The proposed robust models can recover 2.09%, 7.07%, and 35.54% of the lost accuracy for these models. Finally, for the most severe case of attacks affecting 1% of the MRs (with hotspot attacks), the accuracy degradations reach 1.1%, 6.8%, and 13.3% and the robust models are able to recover 1.1%, 6.64%, and 9.07% of the lost accuracy for these models. Thus the best robust models can significantly reduce the impact of HT attacks.

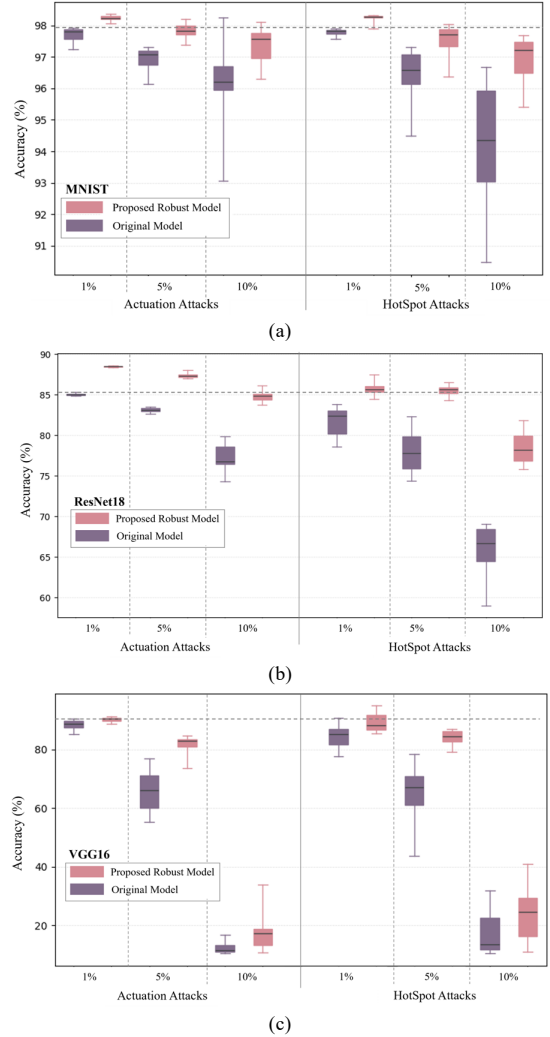

Fig. 9: Accuracy intervals for the proposed robust models and the original CNN models under various actuation and hotspot attacks for: (a) MNIST-based CNN model; (b) ResNet18 CNN model; (c) VGG16 CNN model.

*In summary*, the key insights from our attack mitigation analysis are: *1)* both L2 regularization during training and Gaussian noise aware training can be very effective in compensating for the accuracy drop due to HT-based actuation and hotspot attacks, *2)* combining both strategies provides even more effective attack mitigation, *3)* HTs corrupting 10% of the total MRs in the ONN accelerator can severely damage massive models like VGG16 and the robust techniques explored fail to recover the accuracy drop is such extreme cases, *4)* different configurations of the mitigation strategies provide the most robust outcomes for different CNN models, and *5)* overcoming hotspot attacks is more difficult than actuation attacks.

## VII. CONCLUSION

In this paper, we conducted the first comprehensive susceptibility analysis of actuation and hotspot hardware trojan (HT)-induced attack scenarios and their impact on the performance of multiple CNN models running on a state-of-the-art non-coherent optical neural network (ONN) hardware accelerator. While our results show that HT attacks can significantly degrade model accuracy, we demonstrate that the application of software-based optimization techniques such as L2 regularization and Gaussian noise-aware training can provide a low overhead mechanism to reduce the impact of such attacks. Our ongoing work is exploring how these software techniques can be combined with lightweight hardware-based techniques to provide even more effective HT-induced attack mitigation in ONN platforms.